\def\ts     {\thinspace}
\def\kms    {\ifmmode{{\rm \ts km\ts s}^{-1}}\else{\ts km\ts s$^{-1}$}\fi}
\def\msol   {\ifmmode{{\rm M}_{\odot} }\else{M$_{\odot}$}\fi}
\def\lsol   {\ifmmode{L_{\odot}}\else{$L_{\odot}$}\fi}
\def\lfir   {\ifmmode{L_{\rm FIR}}\else{$L_{\rm FIR}$}\fi}
\def\zsol   {\ifmmode{{\rm Z}_{\odot}}\else{Z$_{\odot}$}\fi}
\def\etal   {{\rm et\ts al.}}
\def\aco    {\ifmmode{{\rm CO}(J\!=\!1\! \to \!0)}\else{{\rm CO}($J$=1$\to$0)}\fi}
\def\bco    {\ifmmode{{\rm CO}(J\!=\!2\! \to \!1)}\else{{\rm CO}($J$=2$\to$1)}\fi}
\def\cco    {\ifmmode{{\rm CO}(J\!=\!3\! \to \!2)}\else{{\rm CO}($J$=3$\to$2)}\fi}
\def\dco    {\ifmmode{{\rm CO}(J\!=\!4\! \to \!3)}\else{{\rm CO}($J$=4$\to$3)}\fi}
\def\eco    {\ifmmode{{\rm CO}(J\!=\!5\! \to \!4)}\else{{\rm CO}($J$=5$\to$4)}\fi}
\def\fco    {\ifmmode{{\rm CO}(J\!=\!6\! \to \!5)}\else{{\rm CO}($J$=6$\to$5)}\fi}
\def\gco    {\ifmmode{{\rm CO}(J\!=\!7\! \to \!6)}\else{{\rm CO}($J$=7$\to$6)}\fi}
\def\hco    {\ifmmode{{\rm CO}(J\!=\!8\! \to \!7)}\else{{\rm CO}($J$=8$\to$7)}\fi}
\def\ico    {\ifmmode{{\rm CO}(J\!=\!9\! \to \!8)}\else{{\rm CO}($J$=9$\to$8)}\fi}
\def\jco    {\ifmmode{{\rm CO}(J\!=\!10\! \to \!9)}\else{{\rm CO}($J$=10$\to$9)}\fi}
\def\kco    {\ifmmode{{\rm CO}(J\!=\!11\! \to \!10)}\else{{\rm CO}($J$=11$\to$10)}\fi}
\def\ci     {\ifmmode{{\rm C}{\rm \small I}}\else{C\ts {\scriptsize I}}\fi}
\def\hi     {\ifmmode{{\rm H}{\rm \small I}}\else{H\ts {\scriptsize I}}\fi}
\def\hh     {\ifmmode{{\rm H}_2}\else{H$_2$}\fi}
\def\cone {\ifmmode{{\rm C}{\rm \small I}(^3\!P_1\!\to^3\!P_0)}
     \else{C\ts {\scriptsize I}{\small$(^3\!P_1\!\to^3\!\!\!P_0)$}}\fi}
\def\ctwo {\ifmmode{{\rm C}{\rm \small I}(^3\!P_2\!\to^3\!P_1)}
     \else{C\ts {\scriptsize I}{\small$(^3\!P_2\!\to^3\!\!\!P_1)$}}\fi}
\def\cij {\ifmmode{{\rm C}{\rm \small I}\,(^3P_i\to^3P_j)}\else{C\ts {\scriptsize I}\,{\small$(^3P_i\to^3P_j)$}}\fi}
\def\cii    {\ifmmode{{\rm C}{\rm \small II}}\else{C\ts {\scriptsize II}}\fi}
\def\tex {\ifmmode{{T}_{\rm ex}}\else{$T_{\rm ex}$}\fi}
\def\tmb {\ifmmode{{T}_{\rm mb}}\else{$T_{\rm mb}$}\fi}
\def\tkin {\ifmmode{{T}_{\rm kin}}\else{$T_{\rm kin}$}\fi}
\def\microns {\ifmmode{\mu{\rm m}}\else{$\mu$m}\fi}
\def\nhh   {\ifmmode{n({\rm H}_2)}\else{$n$(H$_2$)}\fi}
\def\gradv {\ifmmode{{\rm dv/dr}}\else{dv/dr}\fi}
\documentclass[twocolumn]{aa}
\usepackage{graphicx}
\begin{document}
%
 \title{Multiple CO lines in SMM\,J16359+6612 -- Further evidence for a merger
  }

   \author{A. Wei\ss
          \inst{1}
          \inst{,4}
          \and
          D. Downes
          \inst{2}
          \and
          F. Walter
          \inst{3}
          \and
          C. Henkel
          \inst{4}
          }

   \institute{IRAM, Avenida Divina Pastora 7, 18012 Granada, Spain
         \and
             IRAM, 300 rue de la Piscine, 38406 St-Martin-d'H\'eres, France
         \and
             MPIA, K\"onigstuhl 17, 69117 Heidelberg, Germany
         \and
             MPIfR, Auf dem H\"ugel 69, 53121 Bonn, Germany
             }

   \date{}

   \abstract{Using the IRAM 30\,m telescope, we report the detection of the CO(3--2), CO(4--3), CO(5--4) and
     CO(6--5) lines in the gravitational lensed submm galaxy SMM\,J16359+6612 at $z=2.5$. 
      The CO lines have a double peak profile in all transitions.  From a Gaussian decomposition
      of the spectra we show that the CO line ratios, and therefore the 
     underlying physical conditions of the gas, are similar for the 
     blue and the redshifted component.  
     The CO line Spectral Energy Distribution (SED; i.e. flux density
     vs. rotational quantum number) turns over already at the CO 5--4 transition which 
     shows that the molecular gas is less excited than in nearby
     starburst galaxies and high--z QSOs. This difference mainly
     arises from a lower average \hh\ density, which indicates that the gas
     is less centrally concentrated than in nuclear starburst
     regions in local galaxies. We suggest that the bulk
     of the molecular gas in SMM\,J16359+6612 may arise from an
     overlap region of two merging galaxies. The low gas
     density and clear velocity separation may reflect an
      evolutionary stage of the merger event that is in between
      those seen in the Antennae and in the more evolved ultraluminous
      infrared galaxies (ULIRGs) like e.g. Mrk\,231.
   }

   \maketitle
%

\section{Introduction} 

The intensity of the far--infrared (FIR) background indicates that
emission from warm dust contributes significantly to galaxies'  
overall energy output over the Hubble time (Puget \etal\
\cite{puget96}, Fixsen \etal\ \cite{fixsen98}). Recent surveys at 
submm and mm wavelengths with SCUBA at the JCMT and MAMBO at the IRAM
30\,m telescope have identified part of the underlying galaxy population
responsible for the strong FIR emission. Up to now several hundred
submm/mm galaxies (SMGs) have been identified (e.g. Smail \etal\ 
\cite{smail97}, Bertoldi \etal\ \cite{bertoldi00}, Ivison \etal\
\cite{ivison02}, Webb \etal\ \cite{webb03}, Greve \etal\ \cite{greve04}). These surveys,
however, are limited by confusion and therefore 
only trace the bright part of the SMG luminosity function  
(Blain \etal\ \cite{blain02}). Observations of CO, providing
valuable information on the dynamics, size and mass of the
molecular reservoirs in these objects, have only 
been reported for 12 SMGs so far (e.g. Frayer \etal\
\cite{frayer98}, Ivison \etal\ \cite{ivison01}, Downes \& Solomon
\cite{downes03}, Genzel \etal\ \cite{genzel03}, Neri \etal\
\cite{neri03}, Greve \etal\ \cite{greve05}). Due to their selection
based on the mm/submm continuum, all these sources, however, are intrinsically 
luminous, and not representative of the faint end 
of the SMG luminosity function that dominates the FIR background.\\
Recently Kneib \etal\ (\cite{kneib04}) discovered the strongly lensed 
submillimeter galaxy SMM\,J16359+6612 
towards the galaxy cluster A\,2218. This source
is lensed by the cluster into 3 discrete images and the
large magnification (14, 22 \& 9 for images A, B \& C respectively) 
implies that its intrinsic submm flux density 
($S_{850\microns}=0.8$\,mJy, corresponding to $L_{\rm FIR}\approx10^{12}\,\lsol$ (Kneib \etal\ \cite{kneib04}))
is below the confusion limit of existing 850\,\microns\ surveys
($\approx2$\,mJy, Blain \etal\ \cite{blain02}). It therefore 
provides a unique opportunity to investigate a source which is
presumably more representative for the submm population. \\
Detections of CO towards SMM\,J16359+6612 have been reported by Sheth \etal\ 
(\cite{sheth04}, CO 3--2) and Kneib \etal\ (\cite{kneib05}, CO 3--2 \&
7--6). In this letter we report on the detection of the CO 3--2, 4--3,
5--4 \& 6--5 lines towards the strongest lensed component B. 
For the derived quantities in this paper, we use a $\Lambda$ cosmology with $H_{\rm 0} = 71$
\kms\,Mpc$^{-1}$, $\Omega_\Lambda=0.73$ and $\Omega_m=0.27$ (Spergel
\etal\ \cite{spergel03}).\\

\section{Observations}

Observations towards SMM\,J16359+6612 B were made with the IRAM 30\,m
telescope during nine runs between Jan.\ and March 2005 in mostly
excellent weather conditions. We used the AB and CD receiver
configuration with the A/B receivers tuned to the CO(3--2)
($98.310$\,GHz, 3\,mm band) and CO(6--5) ($196.586$\,GHz 1\,mm band) and
C/D to CO(4--3) or CO(5--4) ($131.074, 161.637$\,GHz, 2\,mm band). The beam
sizes/antenna gains for increasing observing frequencies are
25$''$/6.1\,Jy\,K$^{-1}$, 19$''$/6.5\,Jy\,K$^{-1}$, 15$''$/6.9\,Jy\,K$^{-1}$ and 
12.5$''$/7.7\,Jy\,K$^{-1}$. Typical system temperatures were
$\approx$\,120\,K, 220\,K, and 270\,K ($T_{\rm A}^*$) for the 3, 2, and 1\,mm 
observations.  The observations were done in wobbler
switching mode, with switching frequencies of 0.5\,Hz 
and a wobbler throw of $50''$ in azimuth.  Pointing was checked
frequently and was found to be stable to within $3''$. Calibration was
done every 12\,min using the standard hot/cold--load absorber
measurements. The planets Uranus and Neptune were used for absolute
flux calibration.
We estimate the flux density scale to be accurate to about $\pm$10--15\%.\\
Data were taken with the 1\,MHz filter banks on the A/B 3mm
receivers (512 channels, 512\,MHz bandwidth, 1\,MHz channel spacing) and
the 4\,MHz filter banks for the 2\, and 1.3\, mm observations (256
channels, 1\,GHz bandwidth, 4\,MHz channel spacing). The data were processed
with the CLASS software. 
We omitted scans with distorted baselines, and only subtracted linear baselines
from individual spectra.
\begin{figure*}[t] 
\includegraphics[width=18.0cm]{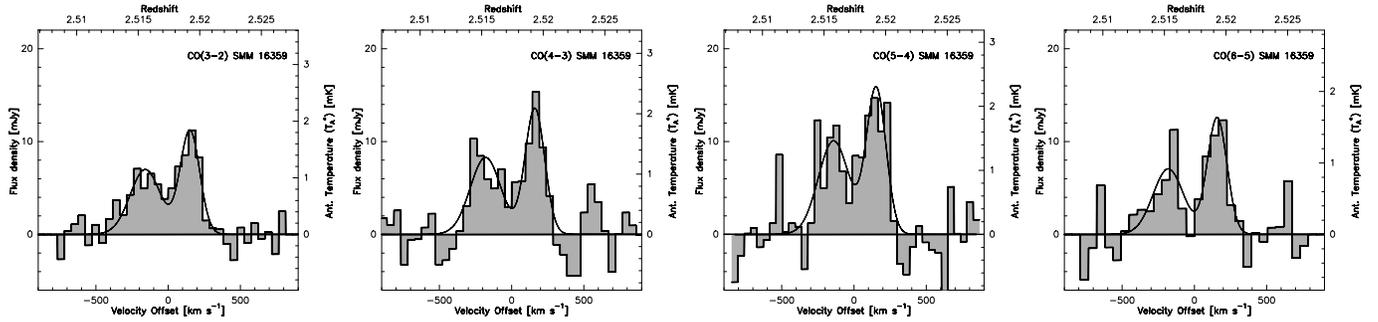}
\caption{Spectra of the CO(3--2), CO(4--3), CO(5--4) and CO(6--5) lines 
 towards SMM\,J16359+6612 B, with Gaussian fit profiles superposed. 
The velocity scale is relative to a CO redshift of
$z=2.5174$. The velocity resolution is 50\,\kms\ (3--2, 4--3, 5--4) 
and  55\,\kms (6--5). All spectra are shown with the same flux scale.}  
\label{lines} 
\end{figure*}
Their average was then regridded to a velocity resolution of 50\,\kms\
(CO 3--2, 4--3, 5--4 ) and 55\,\kms (6--5) leading to rms noise values
($T_{\rm A}^*$) of 0.26\,mK (1.6\,mJy), 0.47\,mK (3.0\,mJy), 0.6\,mK (4.1\,mJy) and
0.38\,mK (2.9\,mJy) respectively.  The total observing time in the
final spectra is 8.7h, 6.8h, 4.7h and 8.2h for the 3--2 to 6--5 lines,
respectively. The final spectra are presented in Fig.~\ref{lines}. We
also generated an averaged CO spectrum by averaging all
individual CO transitions, with equal weight. 
\begin{figure}
\centering
\includegraphics[width=8.5cm]{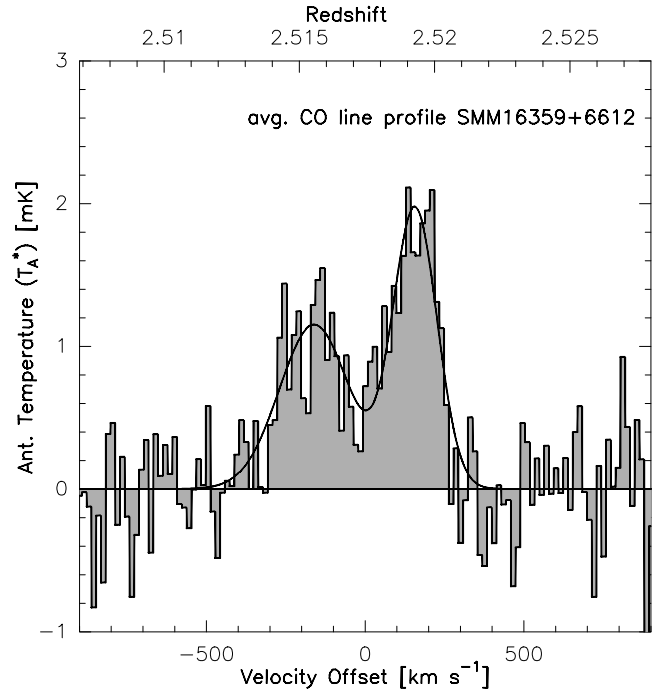} 
\caption{Averaged spectrum of all four CO lines at a velocity
resolution of 15\,\kms\ with Gaussian fit profile superposed. 
All CO lines have equal weight.  The velocity scale is relative to $z=2.5174$. The total
integration time for the spectrum is 28.4 hours (on + off)}  
\label{avg-spectrum} 
\end{figure}
\begin{table}
      \caption[]{Gaussian decomposition of the averaged CO line
         spectrum (see Fig\,\ref{avg-spectrum}). }
         \label{decomp}
          \begin{tabular}{l c c}
            \hline
            \noalign{\smallskip}
            & blue  & red \\
            \noalign{\smallskip}
            \hline
            \noalign{\smallskip}
            
         $z$ & $2.5155\pm 0.0002$ & $2.5192\pm 0.0001$\\
         $\delta$v [\kms]$^{a}$ &$-161\pm17$ &$+156\pm9$ \\ 
         FWHM [\kms] &$250\pm45$ & $160\pm20$\\
         $S_{\nu\,{\rm norm}}$ & $0.58\pm0.18$ & 1\\
         $I_{\rm norm}$ & $0.91\pm0.16$ &  1\\
            \noalign{\smallskip}
            \hline
           \end{tabular}
\begin{list}{}{}
\item[$^{\mathrm{a}}$] Center velocity relative to $z$=2.5174 (Kneib \etal\ \cite{kneib05}).
\end{list}
\end{table}
This spectrum is shown at a velocity
resolution of 15\,\kms\ in Fig.\,\ref{avg-spectrum}.
\section{Results} 
We detected all observed CO transitions from SMM\,J16359+6612 B. 
The line profiles for all four lines are similar, and show the 
characteristic double peak (see Fig.\,\ref{lines}) already 
recognized in previous interferometric studies (Sheth \etal\ 
\cite{sheth04}, Kneib \etal\ \cite{kneib05}). \\
The small separation between the lensed images A and B of $15.0''$
(Kneib \etal\ \cite{kneib04}) implies that our 
spectra also have a contribution from component A. Taking the
relative integrated flux densities between component A/B and 
the beam size of the 30\,m telescope at the four observing frequencies 
into account, we estimate the contribution from component A to the
integrated line intensities of our 
spectra to be 24\%, 12\%, 4\% and 2\% for the 3--2 to 6--5 transitions 
respectively. These corrections have been applied to the following
quantitative analysis. 
Due to the larger distance to component C and its lower magnification, the
contribution from this component is negligible.\\ 
Since the line profiles for all transitions are similar, 
we used the average spectrum of all CO lines
(Fig.\,\ref{avg-spectrum})  to determine the line parameters of the 
blue and redshifted component in the spectrum. A Gaussian
decomposition yields line widths of $250\pm45$\kms\ and 
$160\pm20$\kms\ for the blue and the red components respectively.
The integrated flux density
ratio between the blue and the red component is close to unity
($I_{\rm B}/I_{\rm R} = 0.9\pm 0.2$). The parameters of the spectral 
decomposition are summarized in Table\ \ref{decomp}.\\
For the  Gaussian decomposition of the individual CO lines we fixed
the line width in both components to the values derived above.
The line parameters for each transition are given in Table\ \ref{intlinepara}.
Our integrated CO(3--2) line flux is in agreement with the Bure 
flux reported by Kneib \etal\ (\cite{kneib05}) (2.8 vs 2.5\,Jy\,\kms) 
but lower than the flux reported by Sheth \etal\ (3.5\,Jy\,\kms, \cite{sheth04}). 
   \begin{table*}[t]
      \caption[]{CO line parameters towards SMM\,J16359+6612 B,
      corrected for the contribution from image A. The mean
      redshift is the flux-weighted average of the redshift of the
      red ($R$) and blue ($B$) component. $\delta V_{\rm R,B}$ denotes the
      peak-to-peak velocity separation between the red and blue
      components. Parameters have been
      derived from Gaussian fits by keeping the line widths
      fixed at 250 and 160 \kms\ for the blue and red components.}
         \label{intlinepara}
          \begin{tabular}{l c c c c c c c c}
            \hline
            \noalign{\smallskip}
        Transition & $z_{\rm mean}$ & $S_{\nu\,{\rm R}}$ &
      $S_{\nu\,{\rm B}}$ &
        $\delta V_{\rm R,B}$ & $I$ & $I_{\rm B}/I_{\rm R}$ & $L'\,\,\,^{a}$ & $L\,^{a}$\\
                   & & [mJy]        &[mJy]    & [\kms]     & [Jy \kms]
      & & 
             [10$^{9}$\,K \kms\, pc$^2$] & [10$^{6}\,\lsol$]\\
            \noalign{\smallskip}
            \hline
            \noalign{\smallskip}
            
      \cco\ & 2.51737 & $8.5\pm 0.7$ & $5.3\pm 0.6$ & $310\pm 20$ &
            $2.8\pm 0.2$ & $1.0\pm 0.1$ & $4.4\pm 0.3$ &
            $5.8\pm 0.4$ \\      
      \dco\ & 2.51732 & $11.9\pm 1.5$ & $7.3\pm 1.2$ & $340\pm 25$ &
            $4.0\pm 0.4$ & $1.0\pm 0.2$ & $3.4\pm 0.3$ &
            $10.7\pm 1.1$ \\
      \eco\ & 2.51746 & $15.1\pm 2.0$ & $9.7\pm 1.6$ & $295\pm 30$ &
            $5.1\pm 0.5$ & $1.0\pm 0.2$ & $2.8\pm 0.3$ &
            $17.3\pm 1.8$ \\
      \fco\ & 2.51738 & $12.5\pm 1.9$ & $7.0\pm 1.5$ & $340\pm 35$ &
            $4.0\pm 0.5$ & $0.9\pm 0.2$ & $1.5\pm 0.2$ &
            $16.1\pm 2.1$ \\ 
            \noalign{\smallskip}
            \hline
           \end{tabular}
\begin{list}{}{}
\item[$^a$] corrected for a magnification  $m=22$ (Kneib \etal\ \cite{kneib04})
\end{list}
   \end{table*}
\begin{figure*} 
\centering
\includegraphics[width=17.5cm]{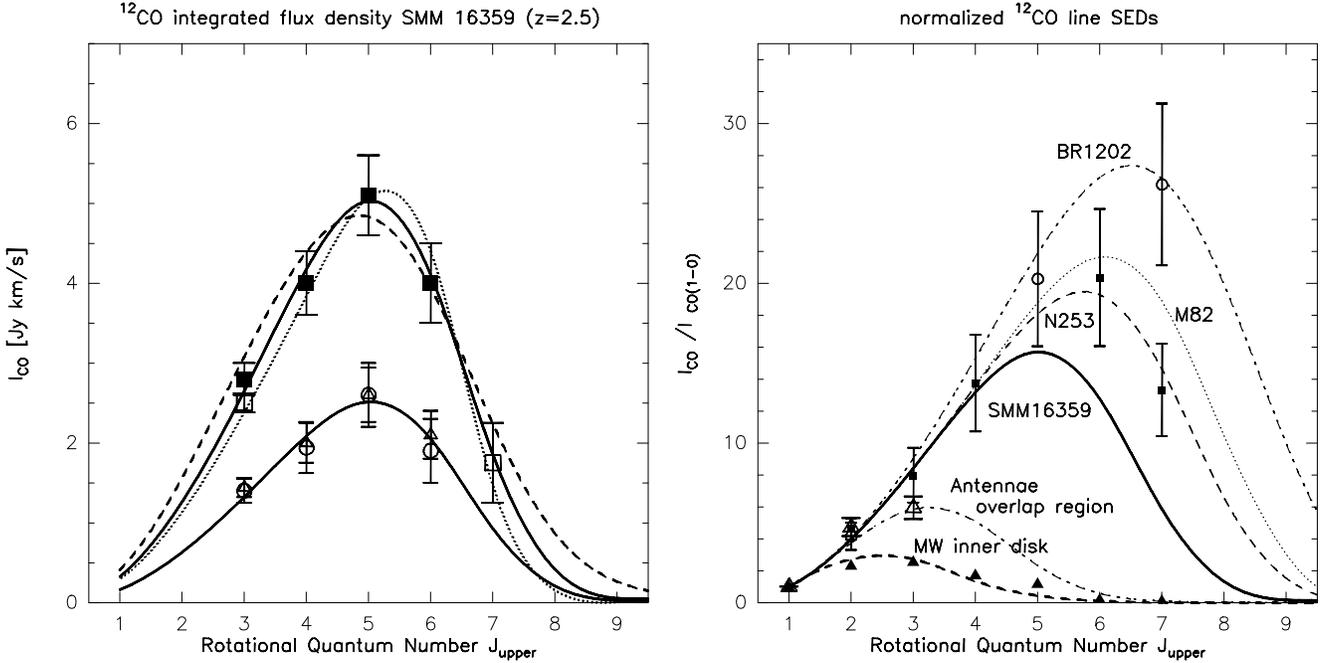} 
\caption{{\it Left:} 
Observed, velocity-integrated CO fluxes vs.\ rotational quantum
number (CO line SED). Filled squares are the integrated flux densities
listed in Table\,\ref{intlinepara}, open squares were adopted
from Kneib \etal\ (\cite{kneib05}). The upper three curves are LVG models
for the entire line profile (red+blue components added together).
The lines show the LVG-predicted fluxes for $n(\hh)$, $\tkin$ of  
$10^{3.4}$ cm$^{-3}$, 35\,K (solid),
$10^{3.1}$ cm$^{-3}$, 80\,K (dashed)
and $10^{3.8}$ cm$^{-3}$, 20\,K (dotted).
The lower data points are the fluxes for the red (open triangles) 
and blue (open circles) components taken separately. The lower curve
shows the same LVG model as the solid upper curve.   
{\it Right:} Comparison of various normalized CO line SEDs: SEDs are
shown for BR1202-0725 ($z=4.7$, open circles Carilli \etal\ \cite{carilli02}), 
the high-excitation component in the center of M82 (Wei\ss\ \etal\ \cite{weiss05}, no data
points plotted), NGC\,253 center (filled squares taken from Bayet \etal\
\cite{bayet04}), SMM\,J16359+6612 (no data points plotted), 
the overlap region in the Antennae galaxies 
(open triangles Zhu \etal\ \cite{zhu03}) 
and the inner disk of the Milky Way (filled
triangles, Fixsen \etal\ \cite{fixsen99}). The CO line SEDs are 
normalized by their CO(1--0) flux density.
} \label{intcosed}
\end{figure*}
The integrated and peak flux densities of the CO lines 
are rising with increasing rotational quantum number up to the 5--4
line.  The line luminosities ($L'$, see e.g. Solomon \etal\
(\cite{solomon97}) for the definition) show that already the 4--3 line
is subthermally excited relative to the 3--2 line. \\ 
To investigate the CO excitation in more detail we used a spherical,
one-component, large velocity gradient model (LVG). We use the
collision rates  from Flower (\cite{flower01}) with an ortho/para \hh\
ratio of 3 and a CO abundance per velocity gradient 
of [CO]/$\gradv = 8\cdot 10^{-5}\,{\rm pc}\,(\kms)^{-1}$. 
All lines are well fitted  with a
{\it single} LVG model.  A good fit to the observations is provided by
n(\hh) = 10$^{3.4}$ cm$^{-3}$ and $\tkin= 35$\,K. Other
temperature--density combinations with similar \hh\ pressure also 
match the data (see Fig.\ \ref{intcosed}). \hh\ densities
above n(\hh)$=\,10^{3.8}$ cm$^{-3}$, however, do not provide a good fit
to the observations.\\ 
In Fig.\,\ref{intcosed} we compare selected
LVG models with the observed integrated flux densities including the
CO 3--2 and 7--6 measurements by Kneib \etal\ (\cite{kneib05}). We
estimated the 7--6 flux using a line ratio of $I_{{\rm
CO}(7-6)}/I_{{\rm CO}(3-2)} = 0.7$ (line integral corrected for the
continuum emission at 1\,mm). The figure shows that the predicted
integrated flux density for the CO(1--0) line is not a strong function of the
model parameters. Our models predict $I_{{\rm
CO}(1-0)} \approx 0.3-0.4$ Jy\,\kms\ or $L'_{{\rm CO}(1-0)} \approx
4.1-5.6\cdot 10^{9}$\,K \kms\, pc$^2$ (corrected for a magnification
$m=22$).\\ 
Our Gaussian decomposition of the individual CO
transitions shows that the integrated flux density ratio between the
blue and the red part of the spectrum is near unity for all CO
transitions (see $I_{\rm B}/I_{\rm R}$ in Table\,\ref{intlinepara}). This implies
that gas masses, CO line ratios and also the underlying excitation
conditions are similar in the red and the blue components.
\section{Discussion}
At first glance the double horn line profile and the similarity 
of the gas properties in both spectral components suggests 
that we are looking at a circumnuclear molecular toroid.  
However, if such a stable molecular gas distribution would supply
the large star formation rate ($\approx500$\,\msol\,yr$^{-1}$ Kneib \etal\ \cite{kneib04})
we would expect much higher gas excitation 
(the same is true if we were looking at independent molecular disks
surrounding the nuclei of two merging galaxies). This is
exemplified in Fig. \ref{intcosed} where we compare the normalized CO line
SEDs (flux density vs. rotational quantum number) of SMM\,J16359+6612 to
those in other starburst galaxies/QSOs. The normalization is relative to
the integrated CO(1--0) flux density (and therefore to the molecular gas
mass). These CO SEDs can be used to constrain the CO excitation as
the turnover of the CO line SED is a sensitive measure of the
underlying gas density and temperature. \\
In SMM\,J16359+6612 the CO SED turnover occurs at the 5--4 line, which
shows that the average gas excitation is lower than in the center of the
well-studied nearby starburst galaxies NGC\,253 and M\,82. In both
cases, the CO SED turns over at the 6--5 line (Bradford \etal\
\cite{bradford03}, Bayet \etal\ \cite{bayet04}, Mao \etal\ \cite{mao00},
Wei\ss\ \etal\ \cite{weiss05}). Even higher CO excitation is
present in Henize\,2-10 (Bayet \etal\ \cite{bayet04})  and some high-z
QSOs such as BR1202-0725 ($z=4.7$, Omont \etal\ \cite{omont96}, Carilli
\etal\ \cite{carilli02}) and the Cloverleaf ($z=2.6$, Barvainis \etal\
\cite{barvainis97}) where the flux density is still rising even up to the
CO(7--6) transition. This comparison shows that the molecular gas pressure
in SMM\,J16359+6612 is lower than in other active systems.  From our
models, we conclude that the difference mainly arises due to the
moderate average gas density (n(\hh) $< 10^4$\,cm$^{-3}$) in
SMM\,J16359+6612. This suggests that the molecular gas is 
less centrally concentrated than gas in starburst nuclei and QSO host
galaxies. This view is supported by the large extent (1.5\,--\,3 kpc) 
of the molecular emission region found by Kneib \etal\ (\cite{kneib05}).\\
One possiblity to explain the combination of lower CO excitation and
high FIR luminosity ($10^{12}\lsol$, Kneib \etal\ \cite{kneib04}) 
may be that the CO flux comes from an 'overlap' region between two 
merging galaxies.
Such a geometry has first been suggested by Kneib \etal\ based on
the UV/optical morphology of the potential two nuclei (their regions
$\alpha$ and $\beta$) and the reddened overlap region (their region
$\gamma$). Such active overlap regions are often found in nearby
interacting systems (e.g., the Antennae: Wilson \etal\
\cite{wilson00}, NGC\,6090: Bryant \& Scoville \cite{bryant99}, Wang
\etal\ \cite{wang04}).\\ 
For comparison, we therefore also show the CO SED for the overlap 
region in the Antennae in Fig.~\ref{intcosed} (Zhu
priv. com.). Although no observations above the CO(3--2) line
have been published in literature so far, the low average 
CO(3--2)/CO(1--0) line ratio suggests that the gas in the 
Antennae overlap region has a lower excitation than
the starburst galaxies/QSOs and SMM J16359+6612.
More detailed models of the molecular gas in the Antennae show
that the bulk of the CO emission arises from gas at moderate density
of n(\hh)$\approx10^{3}$ cm$^{-3}$ (Zhu \etal\ \cite{zhu03}).
Given the early stage of the merging process of the Antennae and its 
low FIR luminosity ($\sim 5\,10^{10}\,\lsol$, Sanders \etal\
\cite{sanders03}) we suggest that SMM J16359+6612 may be a more 
advanced merger (with higher FIR luminosity and higher mean gas 
density than in the Antennae overlap region) but not yet in the 
typical ULIRG stage (with the gas concentrated in the central $<\,1$ kpc
region), such as, e.g., Mrk 231.

\begin{acknowledgements}
We thank M. Zhu for providing us with the CO fluxes for the Antennae
and the IRAM receiver engineers D. John \& S. Navarro as well as
the telescope operators for their great support on optimizing 
the receiver tuning. IRAM is supported by 
 INSU/CNRS (France), MPG (Germany) and IGN (Spain). 
\end{acknowledgements}

\end{document}